\documentclass[12pt]{article}
\usepackage{amssymb}
\usepackage{amsfonts}
\usepackage{amsmath}
\usepackage{amsmath,amsthm}
\usepackage{subfigure}
\usepackage{graphicx,epsfig, color }

\oddsidemargin 10mm \numberwithin{equation}{section}
\def\be{\begin{equation}}
\def\ee{\end{equation}}
\usepackage[numbers,sort&compress]{natbib}
\usepackage[top=3.5cm,right=3cm,bottom=3cm,left=2cm]{geometry}
\begin{document}
\begin{center}
{{\bf {Magnetic charge effects on thermodynamic phase transition
of modified Anti de Sitter  Ay\'{o}n-Beato-Garc\'{i}a  Black Holes
with five parameters}} \vskip 1 cm { Elham Ghasemi
\footnote{E-mail address: e\_ ghasemi@semnan.ac.ir} and Hossein
Ghaffarnejad \footnote{E-mail address: hghafarnejad@semnan.ac.ir}
} }\\
\vskip 0.1 cm
{\textit{Faculty of Physics, Semnan University, P.C. 35131-19111, Semnan, Iran} } \\
\end{center}
\begin{abstract}
In this letter we choose generalized Ayon Beato Gracia (ABG)
magnetic charged black hole with five parameters to investigate
possibility of thermodynamic phase transition and coexistence of
different gass/liquid/solid phases of this black hole. In fact
this work is an extension of our recently work where ABG black
hole with three parameters was used to seek the phase transition.
In this work we obtain other physical values on the parameters
with respect to our previous work where the phase transition is
happened together with coexistence point of different phases in
the phase space.
\end{abstract}
\section{Introduction}
After that Ay\'{o}n, Beato and Garc\'{i}a (ABG) \cite{15}
considered a nonlinear electromagnetic field to produce a
nonsingular magnetic black hole which is applicable to modeling
central black holes of galaxies other authors try to extended his
model with more parameters \cite{16,17,18}. For instance Cai and
Miao \cite{19} take on a generalized ABG black hole solution which
are dependent on five parameters named as the mass, the magnetic
charge and three dimensionless parameters which are related to
nonlinear electrodynamic fields. This kind of black hole returns
to regular black hole under special conditions. In introduction of
our previous work \cite{EG} we describe applications of this type
of the black hole more and we studied its thermodynamics phase
transitions but with some symmetries on parameters of this black
hole. In this work we like to be free of the restrictions on the
parameters of the generalized ABG black hole and we investigate to
study its thermodynamic phase
 transition. In the black hole thermodynamics to produce that equation of state in usual way we need a suitable pressure is affect on the black hole by its environment. This is done by applying a negative cosmological parameter which is originates from anti de sitter vacuum space.
 Fortunately magnetic charge of this kind of black hole can be produce a variable cosmological parameter itself and there is not need to use
  an unknown cosmological constant similar to the well known Schwarzschild de Sitter
  one. In this way, Hawking
and Page discovered a first order phase transition for black holes
in Anti-de Sitter space time \cite{25}. Other types of phase
transitions have been followed in other works
\cite{26,27,28,29,30,31}. Since the cosmological constant has been
suggested as thermodynamic pressure \cite{32,33,34,35}, the
attentions have been attracted to black hole thermodynamics in
extended phase space \cite{36,37,38,39,40,41,42}. Layout of this
work is as follows.\\
In section 2, we define metric of generalized ABG magnetic black
holes with 5 parameters briefly. In section 3, we investigate
thermodynamics perspective of the model. In section 4, we study
possibility of the black hole phase transition and coexistence of
different phases. We dedicate the last section to summary and
conclusion.
\section{ Generalized ABG magnetic black hole}
Consider the following nonlinear Einstein-Maxwell action functional \cite{15}:
\begin{equation}
\label{action} S=\int d^4x \sqrt{-g}
\Big[\frac{R}{16\pi}-\frac{L(P)}{4\pi} \Big]
\end{equation}
in which, $R=g_{\mu\nu}R^{\mu\nu}$ is Ricci scalar and $g=|det
g_{\mu\nu}|$ is absolute value of determinant of metric tensor
field. Nonlinear electromagnetic field lagrangian density
$L(P)=2PH_P-H(P)$ is minimally coupled to the gravity where
$P\equiv\frac{1}{4}P_{\mu\nu}P^{\mu\nu}$ is a gauge invariant
scalar. $P_{\mu\nu}\equiv\frac{F_{\mu\nu}}{H_P}$ is nonlinear
antisymmetric tensor and $F_{\mu\nu}\equiv
\partial_{\mu} A_{\nu}-\partial_{\nu} A_{\mu}$ is electromagnetic tensor field,
where $A_{\mu}$ is electromagnetic potential. $H(P)$ is a
structure function of nonlinear electrodynamic field and
$H_p=\frac{dH(P)}{dP}$ \cite{15}. According to \cite{19}, one can
infer that the above model has a spherically symmetric static
black hole metric field as,
\begin{equation}
ds^2=-f(r)dt^2+f(r)^{-1}dr^2+r^2(d\theta^2+\sin\theta^2d\varphi^2)
\end{equation}
in which,
\begin{equation}
\label{f}
f(r)=1-\frac{2mr^{\frac{\alpha\gamma}{2}-1}}{(q^\gamma+r^\gamma)^{\alpha/2}}+\frac{q^2
r^{\frac{\beta\gamma}{2}-2}}{(q^\gamma+r^\gamma)^{\beta/2}}
\end{equation}
is metric potential with $m$ and $q$ are the mass and the magnetic
charge parameters respectively. Other three dimensionless
parameters $\alpha$, $\beta$ and $\gamma$ are associated to
nonlinear electrodynamic source fields $F_{\mu\nu}$. By assuming
$\alpha\beta\geqslant6$, $\beta\gamma\geqslant8$, and $\gamma>0$
the solution \eqref{f} reduces to regular black hole solutions
\cite{9}. For particular choices of $\alpha=3$, $\beta=4$, and
$\gamma=2$  the generalized ABG metric field \eqref{f} returns to
original ABG black hole solution \cite{15} and it goes to other
generalized ABG black hole solutions \cite{20} by setting
$\gamma=2$. In this work, we consider the metric form (\ref{f}) to
study thermodynamic phase transition of the black hole under
consideration.
\section{Thermodynamic perspective}
In order to participate a negative cosmological  parameter due to
the AdS/CFT correspondence in our study, the metric potential
\eqref{f} can be rewritten similar to Schwarzschild-AdS form
apparently such that
\begin{equation}
\label{ff} f(r)=1-\frac{2M(r)}{r}-\frac{1}{3} \Lambda (r) r^2
\end{equation}
where $M(r)$ is the mass function and $\Lambda (r)$ is the
variable cosmological parameter:
\begin{equation}\label{M1}
M(r)=\frac{m
r^{\frac{\alpha\gamma}{2}}}{(r^\gamma+q^\gamma)^{\alpha/2}},~~~
\Lambda (r)=-\frac{3 q^2 r^{\frac{\beta\gamma}{2}
-4}}{(r^\gamma+q^\gamma)^{\beta/2}}.
\end{equation}
Since the cosmological parameter plays the role of pressure of the
AdS space which influences the black hole, $P(r)$ is defined as
the variable pressure such that
\begin{equation}\label{P1}
P(r)=\frac{-\Lambda (r)}{8\pi}=\frac{3 q^2
r^{\frac{\beta\gamma}{2} -4}}{8\pi (r^\gamma+q^\gamma)^{\beta/2}}
\end{equation}
By rewriting the equation \eqref{ff} in terms of the mass and
pressure parameters and by solving the event horizon equation as
$f(r_+)=0$, we can obtain the enthalpy equation of the black hole
as
\begin{equation}\label{ff1}
M=H=U+PV\end{equation} in which the enthalpy $H$ is equal to the
black hole mass $M$ and the internal energy $U$ and the
thermodynamic volume $V$ are defined respectively by
\begin{equation}U=\frac{r_+}{2},~~~V=\frac{4\pi}{3}r_+^3.
\end{equation}
As we know, the thermodynamic volume is a conjugate quantity of
the thermodynamic pressure and it is different from the geometric
volume of the black hole, but in this case the thermodynamic
volume takes the same form of the geometric volume. By regarding
to the first law of thermodynamic as $dU=TdS-PdV$ and
thermodynamic parameters obtained earlier, we are able to define
the Bekenstein entropy of the black hole under consideration as
follows.
\begin{equation}\label{S1}
S(r_+)=\int \Big[1+8\pi r_+^2 P(r_+)\Big]\frac{dr_+}{2T(r_+)}
\end{equation}
where
\begin{equation}\label{T1}
T_H(r_+)=\frac{f'(r_+)}{4\pi}=\frac{1}{4\pi
r_+}-\frac{q^2r_+^{\frac{\beta\gamma}{2}-2}\left[2(q^\gamma+r_+^\gamma)+(\alpha-\beta)\gamma
q^\gamma\right](q^\gamma+r_+^\gamma)^{-\beta/2}+\alpha\gamma
q^\gamma}{8\pi r_+(q^\gamma+r_+^\gamma)}
\end{equation}
 is the Hawking temperature of the AdS ABG black hole. It is
defined by surface gravity on the black hole horizon $r_+$. In the
latter equation we substituted $m(r_+)$ obtained from the horizon
equation $f(r_+)=0.$ We now investigate possibility of the
thermodynamic phase transition of the AdS ABG black hole.
\section{Equation of state and phase transitions}
By assuming the relation of $x=\frac{r_+}{q}$, the dimensionless
form of mass function and cosmological parameter \eqref{M1} is
obtained as follow:
\begin{equation}\label{m1}
M(x)=\frac{m
x^{\frac{\alpha\gamma}{2}}}{(1+x^\gamma)^{\alpha/2}},~~~
\Lambda(x)=-\frac{3 x^{\frac{\beta\gamma}{2}
-4}}{q^2(1+x^\gamma)^{\beta/2}},
\end{equation}
and the thermodynamic pressure takes the following form:
\begin{equation}\label{p1}
p=q^2P(x)=\frac{3 x^{\frac{\beta\gamma}{2}-4}}
{8\pi(1+x^\gamma)^{\beta/2}}.
\end{equation}
By substituting the relation of $x=r_+/q$ and \eqref{p1} into
\eqref{T1}, the dimensionless equation of state is obtained in
terms of the specific volume $v$ such that
\begin{equation}\label{eq1}
t=qT(x)=pv+\frac{2(1+x^\gamma)+\alpha\gamma}{8\pi x(1+x^\gamma)},
\end{equation}
where
\begin{equation}\label{v}
v=\frac{\left[(\beta-\alpha)\gamma-2\right]x-2x^{\gamma+1}}{3(1+x^\gamma)}.
\end{equation}
As the equation of state reveals the thermodynamic behavior of an
ordinary thermodynamic system and this is valid for black holes
also. In this regard, the equation of state plays an important
role in studying the thermodynamic behavior of a black hole. For
this purpose, \eqref{eq1} is employed to calculate the critical
points within solving the equations of $\frac{\partial t}{\partial
v}\big|_{p}=0$ and $\frac{\partial^2 t}{\partial v^2}\big|_{p}=0$,
which takes the following form due to the usage of the chain rules
to solve the critical equations:
\begin{equation}\label{crit1}
\frac{\partial t}{\partial x}\Big|_{p}=0, ~~~ \frac{\partial^2
t}{\partial x^2}\Big|_{p}=0.
\end{equation}
By substituting (\ref{eq1}) and (\ref{v}) into (\ref{crit1}), one
can obtain parametric forms for the critical points which for
simplicity we substitute ansatz
\begin{equation}\label{chor}x_c=1\end{equation} into them as critical radius
of black hole such that
\begin{equation}p_c=\frac{3}{16\pi}\bigg(\frac{8}{\gamma}-6+2\alpha-\alpha\gamma^2\bigg)\end{equation}
with \begin{equation}\beta=
\frac{\alpha^2(\gamma^4-2\gamma^2)-16\alpha\gamma-32}{\gamma[\alpha(\gamma^3-2\gamma)+6\gamma-8]}.\end{equation}
 By using these conditions into the
temperature (\ref{eq1}) and the volume (\ref{v}), we obtain their
critical values, respectively as follows:
\begin{equation}t_c=\frac{\alpha\gamma^2+2\alpha\gamma+8}{8\pi}\end{equation} and \begin{equation}\label{vc}v_c=-\frac{\gamma}{3}\bigg(\frac{2\alpha\gamma^2+3\alpha\gamma+12}{\alpha\gamma^3-2\alpha\gamma+6\gamma-8
}\bigg).\end{equation}
By substituting (\ref{chor}) into the the horizon equation $f(x)=0$, we obtain
\begin{equation}w=\frac{2m}{q}=2^\frac{\alpha-\beta}{2}(1+2^\frac{\beta}{2}).\end{equation}
To obtain numeric values of the metric parameters of
$(\alpha,\beta,\gamma)$, we keep the positivity condition on the
critical volume $v_c$ and $\gamma>0$ with ansatz
\begin{equation}\alpha=0\end{equation} for which we obtain
 \begin{equation}p_c=\frac{3}{4\pi v_c},~~~t_c=\frac{1}{\pi},~~~\gamma=\frac{4v_c}{2+3v_c},~~~\beta=\frac{(2+3v_c)^2}{2v_c},~~~w=1+2^{-\frac{(2+3v_c)^2}{4v_c}}.\end{equation}
 By using the above numerical values into the equation of
 state (\ref{eq1}) and (\ref{v}) reads
 \begin{equation} \label{eq2}
 v=\frac{2x}{3}\bigg(\frac{1+3v_c-x^\frac{4v_c}{2+3v_c}}{1+x^\frac{4v_c}{2+3v_c}}\bigg),~~~
 t=pv+\frac{1}{4\pi x}.
\end{equation} By substituting $v_c=1$ into
the above equation of state, we plot p-v and t-v diagrams which
are shown in figure 1-a and 1-b. For further investigation on the
phase transition of this ABG magnetic black hole under
consideration, the heat capacity and the Gibbs free energy are
examined. To do so we substitute  $r_+=qx$ into the equation
\eqref{S1} to obtain a dimensionless form for the entropy such
that
\begin{equation}\label{s} ds=\frac{dS(x)}{q^2}=(1+8\pi p
x^2)\frac{dx}{2t}
\end{equation}
which by applying \eqref{eq2} the equation  \eqref{s} reads to the
following series form.
$$
ds=2\pi x-\frac{16 \pi^2 p x^3}{3}+\frac{80 \pi^2 p
x^{19/5}}{3}-\frac{80 \pi^2 p x^{23/5}}{3}+ \frac{512
\pi^3p^2x^{5}}{9}
$$\begin{equation}\label{s1}+\frac{80 \pi^2 p x^{27/5}}{3}-\frac{3200 \pi^3 p^2x^{29/5}}{9}+O(x^6).\end{equation}
By integrating \eqref{s1} we obtain
$$s=\pi x^2-\frac{4\pi^2 px^4}{3}+\frac{50\pi^2px^{24/5}}{9}-\frac{100\pi^2px^{28/5}}{21}+\frac{25 \pi^2 p x^{32/5}}{6}$$
\begin{equation}
+\frac{256 \pi^3 p^2 x^{6}}{27}-\frac{8000 \pi^3
p^2x^{34/5}}{153}+O(x^7).
\end{equation}
Gibbs free energy as other thermodynamic variable is suitable in
study of the phase transition for this modified ABG black hole and
it is obtained from the relation of $G=M-TS$, in which $M$ refers
to the black hole enthalpy or its ADM mass. By substituting the
above equation of state and the entropy equation we obtain a
dimensionless form for the Gibbs free energy such that
$$
g=\frac{G}{q}=\frac{1+2^{-25/4}}{2(1+x^{4/5})}-ts$$$$=
(-86016000\pi^3p^3x^{43/5}+
359661568\pi^3p^3x^{39/5}+6854400\pi^2p^2x^{41/5}-35251200\pi^2p^2x^{37/5}$$
$$+72729600\pi^2p^2x^{33/5}-12343296\pi^2p^2x^{29/5}
-2570400\pi px^{31/5}+367200\pi px^{27/5}$$
$$-489600\pi px^{23/5}-959616\pi px^{19/5}-62390272\pi^3p^3x^7+2924544\pi^2p^2x^5
-5757696\pi px^3$$\begin{equation}-616896x^{9/5}+9639
(2)^{3/4}-616896x+1233792)/(2467584x^{4/5}+2467584).\end{equation}
We plot diagram of the Gibbs free energy versus the temperature at
constant pressures where for pressures higher than the critical
one $p>p_c$ the diagram has a cross point means coexistence of two
phase. Usually this is called as swallowtail form of of the
diagram. This phenomena is appeared also in the diagram of the
Gibbs energy versus the pressure at constant temperatures and
coexistence of two phase is happened at temperatures below the
critical one $t<t_c.$ The Gibbs free energy diagrams are shown in
the figure 2-a and 2-b respectively. We end our study about
possibility of phase transition of modified ABG magnetic black
hole with five parameter by calculating heat capacity of the black
hole. This is obtained as follows.
\begin{equation}
c_p=t\left( \frac{\partial s}{\partial t}\right) _p=\frac{6\pi x^2
(1+8\pi p x^2)(1+x^{4/5})^2}{8\pi p x^2\left[
4-x^{4/5}(1+x^{4/5})\right] -3\left[1+x^{4/5}(2+x^{4/5})\right] }
\end{equation}
where we substitute \eqref{eq2} and \eqref{s}. Change of sign of
the above heat capacity shows phase transition for the black hole
so that positive (negative) heat capacity reveals the absorber
(heater) phase of the black hole. Diagram of the heat capacity at
constant pressure is plotted versus the specific volume. This
shown change of the values of the heat capacity from positive
absorber to negative (heater) values for pressures higher than the
critical one.
\section{Conclusion}
 In this work we used modified nonsingular ABG black hole with five
 parameters to study thermodynamic phase transition by calculating equation of state, Gibbs energy and heat capacity. In fact
 non-singularity of this black hole comes from hypothetical magnetic monopole charge
and this black hole plays important role in modeling of the
Galaxies. Our calculations show that for $P-V$ diagrams the black
hole has three different phases at $t<t_c$ where one phase behaves
as ideal gas with no any phase transition but two other phases
participate in the Hawking Page phase transition. This is happened
because of maximum point of the diagrams where an evaporating AdS
modified ABG black hole reaches to vacuum AdS space finally.
Diagram of T-V at constant pressures shows that the black hole has
three phases for $p<p_c$ where one phase behaves as regular means
that by raising the specific volume the temperature increase. Two
other phases of the black hole participate in a small to large
black hole phase transition because of a minimum point of the
diagram. Coexistence of these phases are studied by plotting the
Gibbs free energy in which crossing point of two branches of the
diagram shows a swallowtail form. Change of sign of the heat
capacity shows positions in phase space where the evaporating
black hole can be a heater/absorber of the thermal energy. This
work with respect to our similar previous work has extensions on
the black hole parameters which causes to phase transition of the
black hole. In our previous work we use some restrictions on the
black hole parameters (with three parameters) corresponding to
model for which Cai and Miao were studied quasi normal modes
\cite{19}. As an extension of this work we like to study effects
of these black hole parameters in the cooling-heating phase or
Jule-Thomson expansion of this black hole.

\begin{figure}[ht]
\centering  \subfigure[{}]{\label{a}
\includegraphics[width=0.4\textwidth]{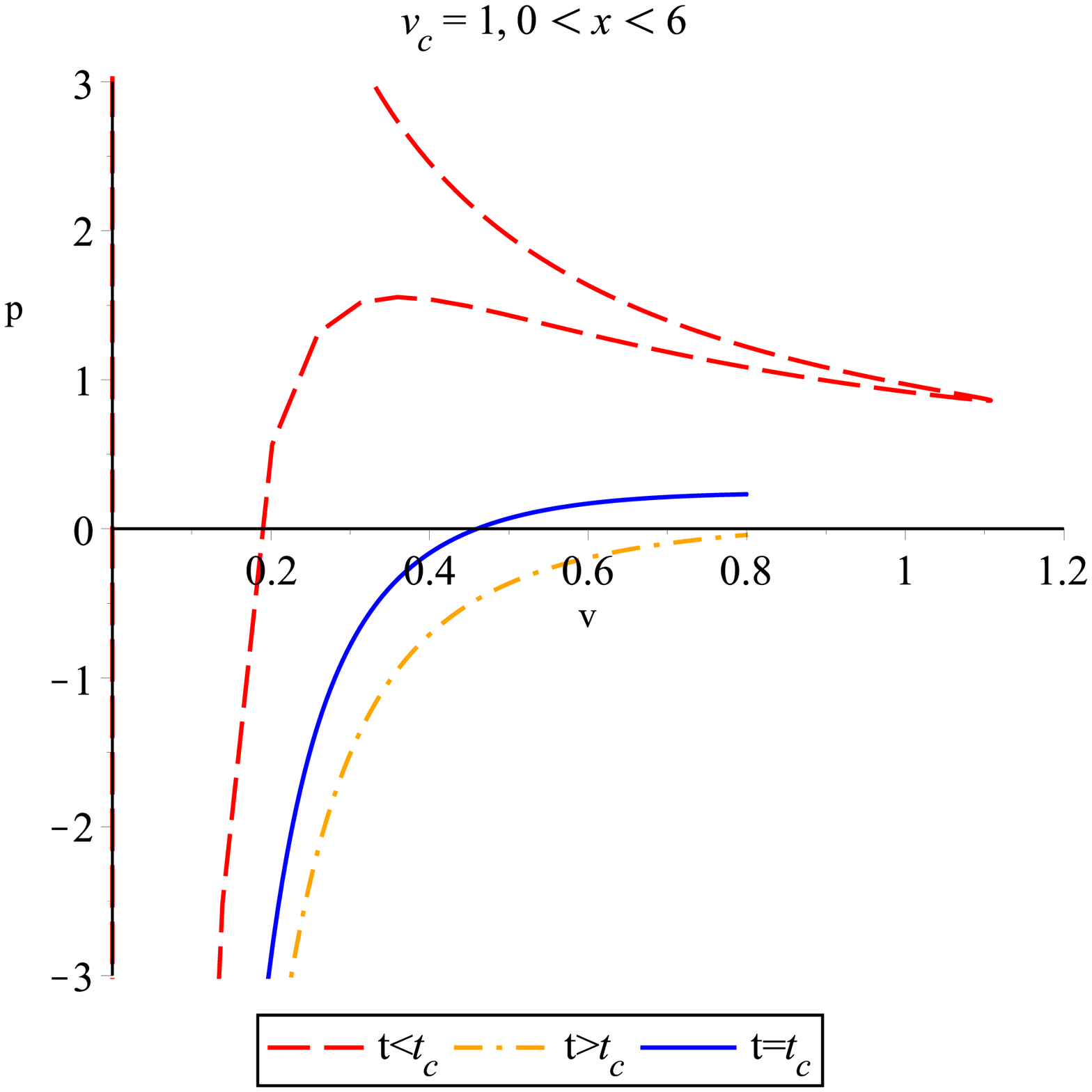}}
\hspace{3mm}\subfigure[{}]{\label{b}
\includegraphics[width=0.4\textwidth]{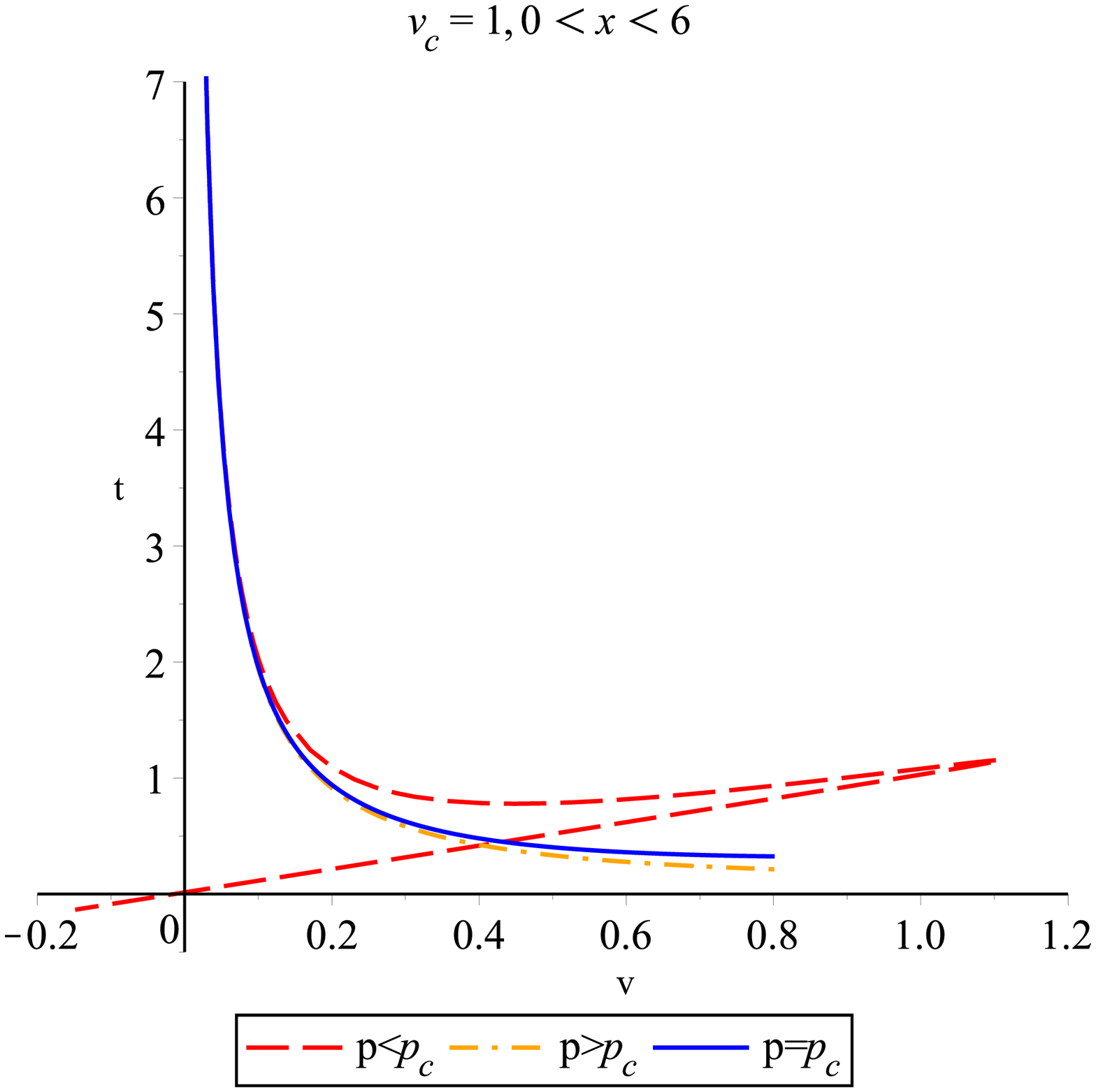}}
\hspace{3mm} \caption{P-V and T-V curves at phase space}
\end{figure}

\begin{figure}[ht]
\centering  \subfigure[{}]{\label{c}
\includegraphics[width=0.4\textwidth]{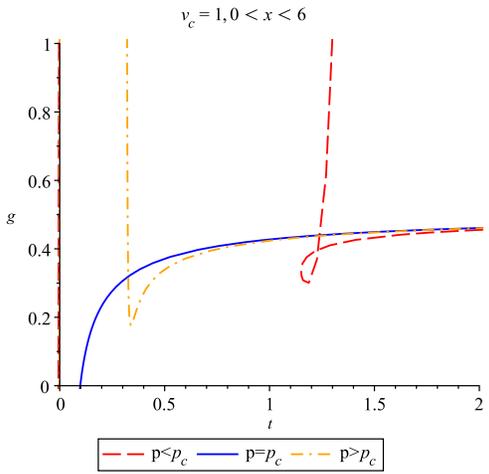}}
\hspace{3mm} \subfigure[{}]{\label{d}
\includegraphics[width=0.4\textwidth]{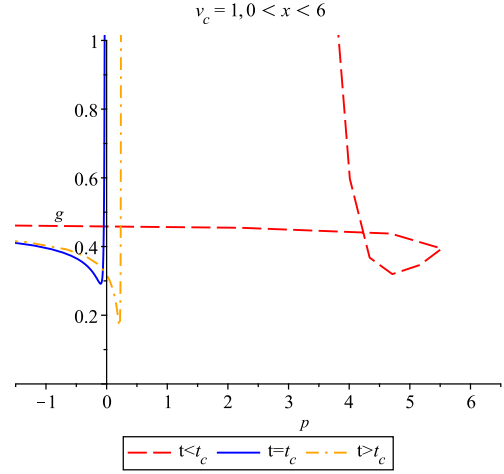}}
\hspace{3mm} \caption{G-T and G-P curves at phase space}
\end{figure}

\begin{figure}[ht]
\centering  \subfigure[{}]{\label{e}
\includegraphics[width=0.4\textwidth]{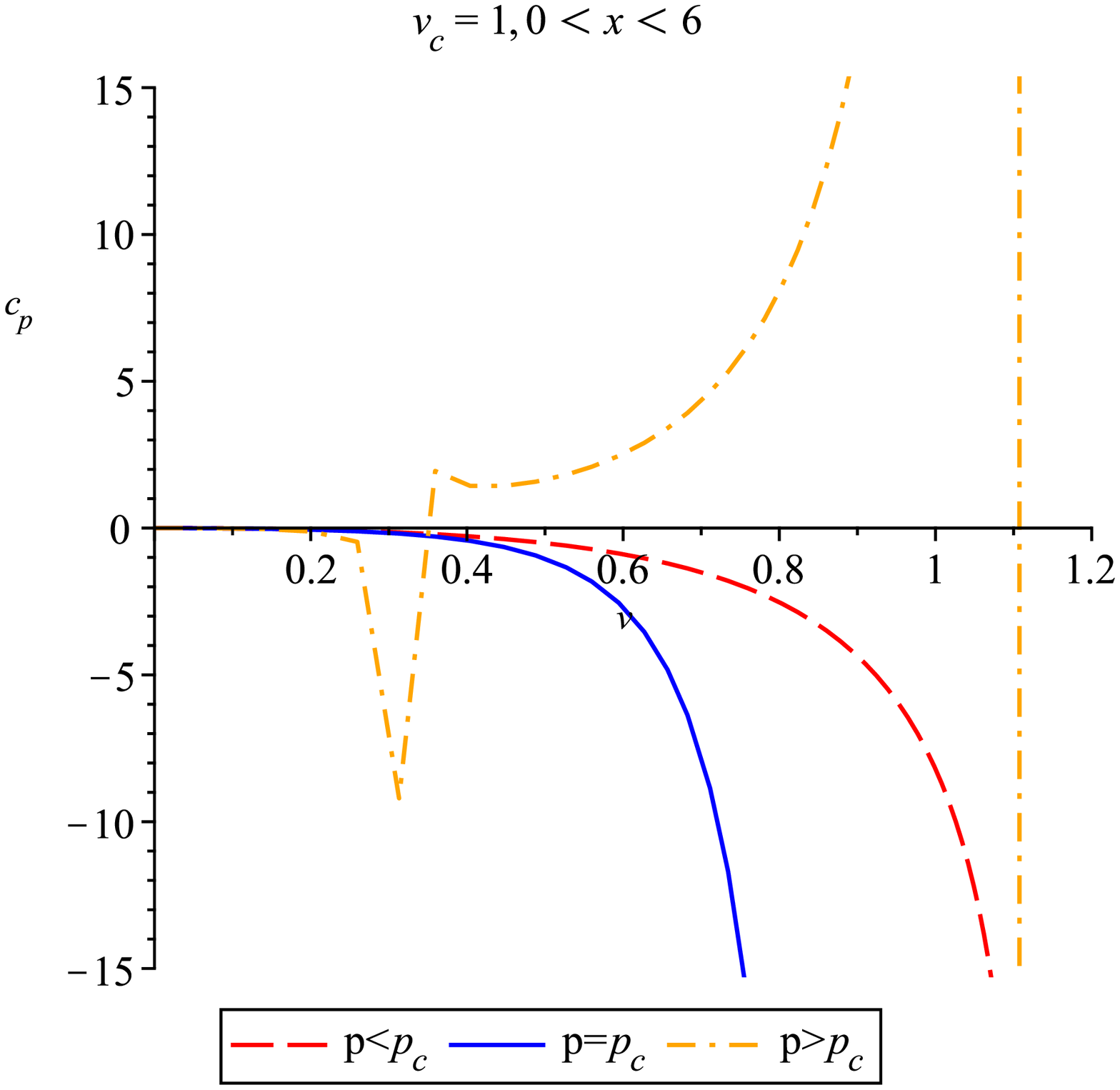}}
\hspace{3mm} \caption{Diagram of heat capacity at constant
pressure versus the specific volume}
\end{figure}


\begin{thebibliography}{99}
\bibitem{15} E. Ayon-Beato and A. Garcia, `Regular black hole in general relativity coupled to nonlinear electrodynamics`, Phys. Rev. Lett. 80 (1998) 5056.
\bibitem{16} L. Balart and E.C. Vagenas, `Regular black holes with a nonlinear electrodynamics source`, Phys. Rev. D 90 (2014) 124045 [arXiv:1408.0306 [gr-qc]].
\bibitem{17} M. S. Ma, `Magnetically charged regular black hole in a model of nonlinear electrodynamics`, Ann. Phys. 362 (2015) 529 [arXiv:1509.05580 [gr-qc]].
\bibitem{18} E.L. Junior, M.E. Rodrigues, and M.J. Houndjo, `Regular black holes in f(T) Gravity through a nonlinear electrodynamics source`, J. Cosmol. Astropart. Phys. 10 (2015) 060 [arXiv:1503.07857 [gr-qc]].
\bibitem{19} X. Ch. Cai and Y. G. Miao, `Quasinormal modes and shadows of a new family of Ay\'{o}n-Beato and Garc\'{i}a  black holes`, Phys. Rev. D 103, 124050 (2021).
\bibitem{20} E. Ghasemi and H. Ghaffarnejad, `Thermodynamic Phase Transition of Generalized Ayon-Beato Garcia Black Holes with Schwarzschild anti de Sitter space time
perspective`, hep-th/2201.08389
\bibitem{25} S. W. Hawking and D. N. Page, `Thermodynamics of black holes in anti-de Sitter space`, Commun. ath. Phys. 87, 577 (1983).
\bibitem{26} Kubiznak, D.; Mann, R.B. `P-V criticality of charged AdS black holes`. J. High Energy Phys. 2012, 1207, 033, doi:10.1007/JHEP07(2012)033.
\bibitem{27} V. P. Maslov,  `Zeroth-order phase transitions`. Math. Notes 2004, 76, 697–710.
\bibitem{28} S. Gunasekaran, D.  Kubiznak, and R. B. Mann, `Extended phase space thermodynamics for charged and rotating black holes and Born-Infeld vacuum polarization`, J. High Energy Phys. 2012, 1211, 110, doi:10.1007/JHEP11(2012)110.
\bibitem{29} N. Altamirano, D.  Kubiznak, and R. B. Mann, `Reentrant phase transitions in rotating antide Sitter black holes`, Phys. Rev. D 2013, 88, 101502, doi:10.1103/PhysRevD.88.101502.
\bibitem{30} N. Altamirano, D. Kubiznak, and R. B. Mann, Sherkatghanad, Z. `Kerr-AdS analogue of tricritical point and solid/liquid/gas phase transition`, ArXiv E-Prints, 2014, arXiv:1308.2672.
\bibitem{31} S. W. Wei and Y. X. Liu, `Triple points and phase diagrams in the extended phase space of charged Gauss-Bonnet black holes in AdS space`, Phys. Rev. D 90, 044057 (2014).
\bibitem{32} D. Kastor, S. Ray and J. Traschen, `Enthalpy and the mechanics of AdS black holes`, Class. Quantum Gravity 26, 195011 (2009)
\bibitem{33} M. Cvetic, G.W. Gibbons and D. Kubiznak, `Black hole enthalpy and an entropy inequality for the thermodynamic volume`, C.N. Pope, Phys. Rev. D 84, 024037 (2011).
\bibitem{34} B. P. Dolan, `The cosmological constant and black-hole thermodynamic potentials`, Class. Quant. Gravity 2011, 28, 125020.
\bibitem{35} B. P. Dolan, `Pressure and volume in the first law of black hole thermodynamics`, Class. Quant. Gravity 2011, 28, 235017.
\bibitem{36} B. P. Dolan, `Where is the PdV Term in the Fist Law of Black Hole Thermodynamics? In Open
Questions in Cosmology`, Olmo, G.J., Ed.; InTech: Rijeka, Croatia,
2012.
\bibitem{37} A. Belhaj, M. Chabab, H. El Moumni, and M. B. Sedra, `On thermodynamics of AdS black
holes in arbitrary dimensions`, Chin. Phys. Lett. 2012, 29,
100401.
\bibitem{38} H. Ghaffarnejad, `Classical and Quantum Reissner-Nordstrom Black Hole Thermodynamics and first order Phase Transition`,
Astrophys. and Space Sci.361,7,6(2016);physics.gen-ph/1308.1323.
\bibitem{39} H. Ghaffarnejad and M. Farsam, `Reissner-Nordstrom black holes statistical ensembles and first order thermodynamic phase transition`, Advances in High Energy Physics, 2019, 2539217 (2019), arXiv:1603.08408[physics.gen-ph].
\bibitem{40} H. Ghaffarnejad, E. Yaraie and M. Farsam, `Thermodynamic phase transition for Quintessence Dyonic Anti de Sitter Black Holes`, Eur. Phys. J. Plus, 135, 179 (2020); arXiv:1808.09789 [physics.gen-ph].
\bibitem{41} H. Ghaffarnejad, `Thermodynamics of 4D dS/AdS Gauss-Bonnet black holes from consistent gravity theory in presence of cloud of
strings`; gr-qc/2105.12729
\bibitem{42} S. W. Wei and Y.X. Liu. `Critical phenomena and thermodynamic geometry of charged Gauss-Bonnet AdS black holes`, Phys. Rev. D 2013, 87, 044014.
\end{thebibliography}
\end{document}